# LOCATING CURRENT SHEETS IN THE SOLAR CORONA

J. BÜCHNER

*Max-Planck-Institut für Sonnensystemforschung, Max-Planck 2, 37191 Katlenburg-Lindau,
Germany (E-mail: buechner@mps.mpg.de)*

**Abstract.** Current sheets are essential for energy dissipation in the solar corona, in particular by enabling magnetic reconnection. Unfortunately, sufficiently thin current sheets cannot be resolved observationally and the theory of their formation is an unresolved issue as well. We consider two predictors of coronal current concentrations, both based on geometrical or even topological properties of a force-free coronal magnetic field. First, there are separatrices related to magnetic nulls. Through separatrices the magnetic connectivity changes discontinuously. Coronal magnetic nulls are, however, very rare. Second, inspired by the concept of generalized magnetic reconnection without nulls, quasi-separatrix layers (QSL) were suggested. Through QSL the magnetic connectivity changes continuously, though strongly. The strength of the connectivity change can be quantified by measuring the squashing of the flux tubes which connect the magnetically conjugated photospheres.

We verify the QSL and separatrix concepts by comparing the sites of magnetic nulls and enhanced squashing with the location of current concentrations in the corona. Due to the known difficulties of their direct observation, we simulated coronal current sheets by numerically calculating the response of the corona to energy input from the photosphere, heating a simultaneously observed Extreme Ultraviolet Bright Point. We did not find coronal current sheets at separatrices but at several QSL locations. The reason is that, although the geometrical properties of force-free extrapolated magnetic fields can indeed hint at possible current concentrations, a necessary condition for current sheet formation is the local energy input into the corona.

## 1. Introduction

The million degrees hot solar corona, flares, and coronal mass ejections (CME), the permanent release of magnetic energy, – all are due to dissipation of energy transported from the center of the sun through the photosphere to the solar atmosphere. In the hot and dilute corona magnetic energy can be dissipated either due to wave damping or due to magnetic reconnection. Though suggested already 60 years ago by Giovanelli (1946), the physics of reconnection is still under investigation. In particular, it is still unclear, where in the corona plasma instabilities can enhance collisionless dissipation to allow reconnection. Current sheets, subject to further filamentation, are potential sites, where this can happen.

Different theories predict the formation of current sheets in the corona (see, e.g., Parker, 1994). Observational methods allow a direct investigation only of thick current sheets, evolving in the aftermath of CMEs (Bemporad *et al.*, 2006), to estimate their thickness and diffusivity. Usually, the noise broadening of the atomic and ionic lines inhibits chromospheric and coronal magnetic field measurements (Solanki *et al.*, 2003). Hence, theoretical investigations are necessary to derive the properties of current sheets and their location in the solar atmosphere.

To find their sites, one can ask for the constraints which the reconnection theory imposes on the location of current sheets. In the limiting case of two dimensions (2D), i.e., for a translational invariance in the third dimension with no magnetic field component pointing in this direction, Vasyliunas (1975) defined reconnection by the flow of a magnetized plasma through separatrices. In this artificial limit, the separatrices intersect at a straight separator line along which the magnetic field vanishes. Separatrices divide the space into regions of different magnetic field topology and host the current sheets. In three dimensions, there is no such obvious location of current sheets. In analogy to the separator – a null line in 2D – in three dimensions (3D) null points might be important for reconnection (Bulanov and Olshanetskii, 1984; Craig and McClymont, 1991; Hassam, 1992; Titov and Priest, 1993). Magnetic nulls can be found, indeed, in the corona above active regions (see, e.g. Aulanier *et al.*, 2000). Numerical simulations also support the idea that 3D magnetic nulls may play a role in the relaxation of the magnetic tension in the corona (Antiochos *et al.*, 2002). In analogy to the 2D limit, the formation of current sheets is expected near 3D separators, lines, which connect coronal magnetic nulls (see, e.g., Somov and Titov, 1985; Gorbachev and Somov, 1988; Longcope and Cowley, 1996). Separatrices in 3D are formed by the surface of the fan field lines of a magnetic null connected to another null. Three-dimensional separatrices divide regions of different magnetic field topology. As in the 2D limit, they intersect at separators (see, e.g., Priest and Forbes, 2000). Separators exist not only between nulls but also between "bald patches." These are regions of the photosphere, where the surface-normal magnetic field vanishes (Titov *et al.*, 1993). In order to find out, whether at separatrices current concentrations arise, Zwingmann *et al.* (1985) imposed a shear flow through the photospheric footpoints of separatrices in a two-and-a-half dimensional configuration, translational invariant in the third dimension with an added constant guide magnetic field in the invariant direction. They found that, indeed, current sheets form along the separatrices. Later Schindler *et al.* (1988) considered the case of a flux tube rotation. They concluded that in the transition to 3D separatrices might become structurally unstable and disappear. Aly (1990) and Lau and Finn (1992) found current sheets along the separatrices of 3D magnetic field configurations without translational invariance. And, finally, Démoulin *et al.* (1996b) could demonstrate for a quadrupolar magnetic field model that separatrices do not necessarily become structurally unstable in the transition to 3D.

Since separatrices need 3D magnetic nulls or bald patches, the question arises, how common coronal magnetic nulls are? In order to simplify the description of the complex coronal magnetic field, different methods were developed including that of reducing the distributed sources to magnetic monopoles located either below or in the photosphere (for details see, e.g., Longcope and Klapper, 2002; Priest *et al.*, 2005 and references therein). This method, however, overestimates the occurrence of nulls, it reveals additional nulls which disappear if one returns to the original, distributed photospheric magnetic field. Nevertheless, even for coronal field models generated by magnetic monopoles, only a small number of magnetic nulls can be found in the corona (see, e.g., Schrijver and Title, 2002; Longcope *et al.*, 2003 and references therin). In more realistic models, considering distributed photospheric magnetic field sources, the number of coronal nulls further decreases.

If, therefore, coronal magnetic nulls and, therefore, genuine separatrices, are not very common, the question arises, where else can currents become concentrated? 3D reconnection does not require magnetic nulls (Hesse and Schindler, 1988). To start, again, with the theory of reconnection, note that reconnection can take place also without nulls. Hence, current sheets may be formed without separatrices as well? Indeed, Priest and Demoulin (1995) have demonstrated that reconnection might take place if just the magnetic linkage (connectivity) changes strong enough instead of discontinuously jumping like at separatrices. In analogy to separatrices, they called such regions Quasi-Separatrix Layers (QSLs) (see, e.g., Démoulin *et al.*, 1996a,b). Priest and Demoulin (1995) suggested to quantify the strength of the connectivity change by calculating the norm $N = ||J||$ of the Jacobian matrix $J$ that relates the shift-vector of a photospheric footpoint motion (independent variable) to the shift-vector of the footpoint motion of the field line in the magnetically conjugated photosphere (for details see Section 2, a comprehensive review of the generalization of separatrices to QSLs can be found e.g., in Démoulin (2005, 2006). Note that the QSL concept is in accordance with the most general definition of reconnection as a magnetic connectivity change through a nonideal plasma region (Axford, 1984). The collisionless coronal plasma can become nonideal if the current densities let the current carrier velocity exceed an appropriate plasma instability threshold.

How well can coronal current sheets be predicted by QSL (see, e.g., Demoulin *et al.*, 1997)? Due to the complex structure of the coronal magnetic field and the lack of direct observations of current concentrations it is appropriate to utilize numerical simulation techniques to investigate this question. For example, Aulanier *et al.* (2005) simulated the formation of coronal current sheets in a, strictly speaking, bipolar configuration, modelled by four magnetic flux concentrations. In such a configuration, the absence of a coronal magnetic null is guaranteed. After the authors imposed a smooth, sub-Alfvénic photospheric footpoint motion, extended current sheets evolved. Narrow current layers developed all around the QSLs, determined for the initial potential field extrapolation. The strongest currents were created in places where the QSL were the thinnest, at so-called Hyperbolic Flux Tubes (HFT), which generalize the concept of a separator (Titov *et al.*, 2002). If the QSL were broader than the dissipative length scale of the code, the currents around the QSL gradually grew to large amplitudes.

Büchner (2005) used the simulation of chromosphere and corona simulation of an observed Extreme Ultraviolet (EUV) Bright Point (BP) to determine, starting with the observed photospheric magnetic fields and plasma motion, the sites of parallel reconnection electric fields in a configuration without nulls in the corona above. They compared the reconnection sites with the position of chromospheric magnetic nulls and enhanced differential flux tube volumes. In their approach resistivity is switched on, if the current carrier velocity $j_{||}/\rho$ (where $j_{||}$ is the parallel current density and $\rho$ the local charge density) exceeds a certain plasma-physically determined threshold. They found that the parallel current carrier velocity is enhanced in regions of enhanced differential flux tube volumes but away from the chromospheric magnetic nulls.

Here, in this paper, we extend our previous work to the comparison of the location of parallel and perpendicular, non-force-free, currents with QSL. First, in Section 2, we review two versions of a squashing factor (Titov *et al.*, 2002) that quantifies the strength of the magnetic connectivity change in a symmetric way. In Section 3 we present the results of a simulation of current concentrations for the observed distributed photospheric magnetic fields. Energy is added to the corona by rotating one of the magnetic flux concentrations as observed. To locate the current concentrations and the QSL, we map the parallel and perpendicular (to the coronal magnetic field) current concentrations to the photosphere by integrating the current densities along the magnetic field and compare them with the location of enhanced squashing factors. Finally, in Section 4, we draw our conclusions about the relation between regions of enhanced squashing indicating strong or even discontinuous changes of connectivity in the potential magnetic field configuration, extrapolated out of the observed longitudinal photospheric field, and the locations of current concentrations in the corona perturbed by energy inflow.

## 2. Locating Quasi-Seperatrix Layers

One of the hypotheses about the creation of current concentrations in the solar corona in the absence of magnetic nulls and separatrices is concerned with the presence of regions of strong gradients in the magnetic connectivity. Priest and Demoulin (1995) suggested to quantify the magnetic connectivity by the norm $N_\pm = ||J_\pm||$ of the Jacobian matrices of the vectors functions describing the shift of photospheric footpoints of magnetic field lines in their dependence on the shift of the footpoint of this field line in the magnetically conjugate photosphere. The sign + stands for a forward mapping $(x_+, y_+) \to (x_-, y_-)$ and − for the reverse one $(x_-, y_-) \to (x_+, y_+)$, respectively. The shift vectors are given by $\{X_-(x_+, y_+); Y_-(x_+, y_+)\}$ and $\{X_+(x_-, y_-); Y_+(x_-, y_-)\}$ in Cartesian coordinates. The Jacobian matrices $J$ describe the shift of the magnetically conjugate footpoint of the field line with respect to a shift of the starting footpoint. Let us denote the starting point of a forward transformation by a + sign and the backward transformation by $-$. Then the resulting two norms are $N_+ = N(x_+, y_+)$ and $N_- = N(x_-, y_-)$, given by

$$N_\pm = \sqrt{\left(\frac{\partial X_\pm}{\partial x_\pm}\right)^2 + \left(\frac{\partial X_\pm}{\partial y_\pm}\right)^2 + \left(\frac{\partial X_\mp}{\partial x_\pm}\right)^2 + \left(\frac{\partial X_\mp}{\partial y_\pm}\right)^2} \quad (1)$$

Démoulin *et al.* (1996b) defined QSL as regions, for which the norms $N_\pm$ become large. $N_\pm \gg 1$ indicate regions of strong change of the magnetic connectivity. Generally the normal magnetic field components at the conjugated ends of field lines are different. Hence, the two norms $N_+$ and $N_-$, calculated for the same field line but starting from the one and the other photospheric footpoints, are different as well. To avoid an ambiguous characterization of the same magnetic link, Titov *et al.* (2002) suggested to multiply the norms calculated for the two ends of a field line and to normalize their product by the determinant of the Jacobian. Since the geometrical analogy is the eccentricity of an elliptical cross-section of the flux tube into which a flux tube with an initially circular cross-section is transformed, Titov *et al.* (2002) called the resulting quantity "squashing factor," given by

$$Q_1 = \frac{N_+^2}{\det ||J_+||} = \frac{a^2 + b^2 + c^2 + d^2}{|ad - bc|} \quad (2)$$

where

$$a = \left(\frac{\partial X_-}{\partial x_+}\right), \quad b = \left(\frac{\partial X_-}{\partial y_+}\right), \quad c = \left(\frac{\partial Y_-}{\partial x_+}\right), \quad d = \left(\frac{\partial Y_-}{\partial y_+}\right) \quad (3)$$

Titov *et al.* (2002) redefined a QSL as a region with $Q_1 \gg 2$.

The determinant in the denominator of $Q_1$ contains the derivatives $a \ldots d$ given by Equation (3). It can be replaced by the ratio of the cross section of the flux tube at the conjugate footpoints, $S_e/S_s$. By definition of a magnetic flux tube, the ratio of the cross-sections can be replaced by $|Bzs/Bze|$, the ratio of the normal to photosphere magnetic field components at the two ends of the field line. Note that the direction of the mapping has to be taken into account. As suggested by Titov *et al.* (2002), this simplifies the calculation of the squashing factor to

$$Q \equiv Q_+ = \frac{N_+^2}{|B_{zs+}/B*_{ze-}|} \equiv Q_-^* = \frac{N_-^2}{|B*_{zs-}/B_{ze+}|} \quad (4)$$

where the asterisks indicate that the arguments $x_-$ and $y_-$ are replaced by the ones of the magnetically conjugate footpoints $x_+$ and $y_+$.

In order to estimate $Q$ and $Q_1$, one can implement, e.g., the following algorithm: For any starting position of interest $\{x_{s0}, y_{s0}, 0\}$ at the photosphere (hence $z_{s0} = 0$) one determines the coordinates of the magnetically conjugate photospheric footpoint of the field line $\{x_{e0}, y_{e0}, 0\}$. To estimate the squashing of an initially circular flux tube to an ellipse, it is sufficient to calculate two additional magnetic mappings starting at footpoints 90° apart along the perimeter of a circle with a radius $\Delta x = \Delta y = r$ around $\{x_{s0}, y_{s0}, 0\}$. Now one is required to find the conjugate footpoints for the second and third field lines, starting, e.g., at $\{x_{s1}, y_{s1}, 0\} = \{x_{s0}+\Delta x, y_{s0}, 0\}$ and $\{x_{s2}, y_{s2}, 0\} = \{x_{s0}, y_{s0}+\Delta y, 0\}$, respectively. If one denotes the resulting conjugate footpoints of these field lines $\{x_{e1}, y_{e1}, 0\}$ and $\{x_{e2}, y_{e2}, 0\}$, their deviations from $\{x_{e0}, y_{e0}, 0\}$ provides a good estimate for the elements of the Jacobian matrix:

$$a \cong \frac{x_{e1} - x_{e0}}{\Delta x}, \quad b \cong \frac{x_{e2} - x_{e0}}{\Delta y}, \quad c \cong \frac{y_{e1} - y_{e0}}{\Delta x}, \quad d \cong \frac{y_{e2} - y_{e0}}{\Delta y} \quad (5)$$

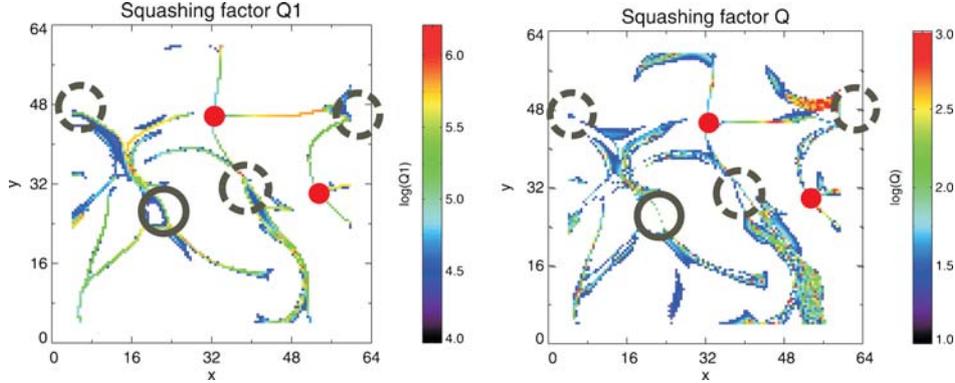

*Figure 1. Left panel:* Squashing factor $Q1$ for the exact determinant in the denominator (cf. Equation (2)). *Right panel:* Squashing factor $Q$ for the simplified by the ratio of the normal magnetic field components denominator (cf. Equation (4)). The *color* coding depicts the logarithm of the squashing factors. The *red dots* indicate the positions of the two chromospheric magnetic nulls located just above the photosphere. The *black solid* and *dashed circles* show the approximate size and location of the main magnetic flux concentrations in the field of view, the same line style means the same polarity.

A sensitive choice is that of $r = \Delta x = \Delta y$. It must be chosen not too large in order to allow the detection of possibly thin layers in case of strong connectivity change, like at HFT, near separatrices. On the other hand $r$ should not be chosen too small, e.g., not smaller than the finite resolution of the magnetic field. The calculation of too many $Q$'s to cover the field densely is computationally very expensive. Following this procedure, we determined the squashing factors $Q$ and $Q_1$ for the chosen magnetic field configuration, for which we simulated the coronal currents (see Section 3).

Figure 1 depicts the two versions of the squashing factor obtained in the area of interest. We use a logarithmic scale since the squashing factors become very large inside a QSL. The left panel shows $Q_1$ with the determinant of the Jacobian containing the derivatives in the denominator (cf. Equation (2)). The right panel shows $Q$, calculated with the ratio of the normal magnetic field components in the denominator (cf. Equation (4)). The red dots in the plots indicate the positions of two magnetic nulls which, as shown by Büchner (2005), are located very low in the chromosphere at {33; 46; 0.5} and at {54; 30; 1} in the normalized to one MDI pixel {$X$; $Y$; $Z$} coordinates. Hence, with their $Z$-coordinates of 0.5 and 1 in normalized units, their height above the photosphere is about 230 and 460 km, respectively, in absolute units. The solid and dashed black circles in Figure 1 indicate the approximate size and location of the main photospheric magnetic flux concentrations in our area (same line styles mean same polarity). As one can see, in Figure 1, the peak values of the squashing factor $Q_1$, calculated for the determinant in the denominator, exceed the $Q$ values, obtained with the approximated denominators, by three orders of magnitude. $Q_1$ also provides a sharper contrast than $Q$, i.e., $Q_1$ reveals thinner QSL, if the width of a QSL is determined as the width of the $Q$ distribution at half of the peak value. Obviously, both $Q_1$ and $Q$ are maximum at separatrices, where the magnetic connectivity changes not just strongly as at QSL, but discontinuously. The photospheric footpoints of the separatrices, the lines of maximum squashing, can easily be found in Figure 1, where the positions of the chromospheric nulls are indicated by red dots. Smaller, but still large, squashing factors can be found away from the two chromospheric nulls and away from the separatrices through them. They indicate real QSL and their photospheric footpoints are located inside the magnetic flux concentrations rather than near nulls.

## 3. Locating Current Concentrations

Let us now locate the coronal currents to compare them with the location of the QSL, found in the previous section. Coronal currents are created in response to the energy input from below the photosphere. As well known for complex dissipative systems, the solar atmosphere reacts to the energy inflow by structure formation, creating currents in the corona. We choose a well-observed EUV-BP situation to find the currents by numerical simulation. EUV BPs are local brightenings in the coronal EUV radiation. The energization of BPs is usually referred to as being related to local magnetic reconnection. Traditionally, BP-related reconnection scenarios rely on the formation of separatrices (Parnell *et al.*, 1994; Priest *et al.*, 1994) or separators (Longcope, 1998) in the coronal magnetic field. Indeed, BPs are often found between moving oppositely directed magnetic flux concentrations. As the magnetic polarities move apart, the interconnecting magnetic flux can reconnect, e.g., to the overlying magnetic field. For a motion of the polarities toward each other reconnection through null points and separatrices can take place. But coronal magnetic nulls are rare (cf. Section 1) while BPs are often observed. Also, as the conjugate magnetic fluxes below BPs are not always moving at or away from each other, rotations and cancellation of fluxes were observed in BP regions as well. We simulated the location of coronal currents, starting with the extrapolated magnetic field, based on photospheric observations. The energy input through the photosphere is incorporated by considering the observed photospheric plasma motion as a boundary condition (Büchner *et al.*, 2004a,b). Note that the magnetic field extrapolation method has to be carefully chosen, in order to provide MagnetoHydroDynamics (MHD)-compatible boundary conditions (Otto *et al.*, 2006). After adding an equilibrium solar plasma and neutral gas filling, the neutral gas is moved in accordance with the observed photospheric plasma motion. Due to the strong plasma-neutral gas coupling in the chromosphere, the plasma follows the photospheric motion, carrying the frozen-in magnetic field. This perturbs the coronal plasma and magnetic field and creates currents. In the past, we had considered the cases of two magnetic polarities moving at each other (Büchner *et al.*, 2004b) and of such, moving apart (Büchner *et al.*, 2005). Here, in this paper, we use the results obtained for a rotating magnetic flux concentration observed during the Π-phase of a BP on June 14, 1998, starting at 12:00 UT (Brown *et al.*, 2001). We simulate the corona above an 28 Mm × 28 Mm wide area close to the central meridian and to the solar equator, the same as considered in Section 2. We choose this area, since it contains the two main flux concentrations as well as several smaller side-polarities. The latter can, according to Podladchikova and Büchner (2006), essentially contribute to the energization of an EUV-BP. The Π-phase of this BP evolution was characterized by a counterclockwise rotation of the large unipolar magnetic flux concentration located at a distance of $\{x, y\} = \{25, 25\}$ MDI pixels (1 pixel = 463 km in the photosphere) from the southeastern vertex of the area. After the energy input is started, a major parallel current sheet is formed from the chromosphere to the corona through the transition region. In addition to the parallel currents, arising first, also currents flowing perpendicular to the ambient magnetic field are created, for which $[\mathbf{j}_\perp \times \mathbf{B}] \neq 0$ (cf. Büchner and Nikutowski, 2005). In order to compare the arising current concentrations with the location of QSL, we separately integrated both the parallel and perpendicular current densities along the magnetic field. The left panel of Figure 2 relates the integrated parallel, the right panel the integrated perpendicular current densities, to the photospheric footpoints of the field lines along which the integration was carried out. The total perpendicular current density integrated along the magnetic field line is much smaller than the field-line-integrated parallel current density, as one can see comparing the two panels of Figure 2 (since the length of the field lines along which the current densities are integrated is the same for both). Also, though weaker (peak values: 17 simulation units or $\int j \, dl = 5 \times 10^4$), return currents are formed to the sides of the main current concentrations which reach peak values of 34 simulation units or $\int j \, dl = 5 \times 10^5$. Note that in order to reach a current carrier drift velocity of the order of the electron thermal velocity $v_{\text{te}} = 4 \times 10^6$ m/s for coronal temperatures ($T_e = 10^6$ K), the current densities in a $n = 10^{16}$ m$^{-3}$ dense coronal plasma should be $6 \times 10^2$ A/m$^2$. Such high current density can be reached if the integrated current density of $\int j \, dl = 5 \times 10^5$ is equally distributed along the whole flux tube in a 1 km, i.e., a proton gyroradius thick current sheet. Since the current carrier density becomes maximum locally, near the transition region, the sheets can be even thicker – though it will be still below the spatial resolution of telescopes and numerical fluid codes (for details, see Büchner *et al.*, 2005). Both, the parallel as well as the (non-force-free) perpendicular currents are enhanced mainly around the rotated main flux concentration, through which most of the energy enters the chromosphere (left central maximum, encircled by the solid circles). Strong currents are generated also around the conjugated second main magnetic polarity (encircled by a dashed line), to which most of the flux from the first main polarity connects. In addition, one finds current concentrations along a bowing line, which starts from the second main polarity. The reason is that, in the course of its rotation, the main (left) polarity changes its linkage to the magnetically conjugate photosphere to locations along this line. As we have shown in the previous Section 2, the bow-shaped line corresponds to the footpoints of a QSL (see the plotted squashing factors in Figure 1). The currents along the bowing line starting from the second main polarity follow pretty much the brightening observed by TRACE (Brown *et al.*, 2001; Büchner *et al.*, 2004a). Hence, it may well be that the observed EUV brightening is related to dissipation of these currents, either directly by Ohmic heating or via reconnection. If one assumes a certain threshold at which anomalous resistivity switches on, one can find the location of parallel electric fields which indicates finite-B-reconnection. With the assumption of a threshold in the parallel current carrier velocity $j_{\|}/\rho$ one finds that reconnection takes place in the the transition region, far away from the magnetic nulls (Büchner *et al.*, 2004a). Comparing Figure 1 with Figure 2, one sees that the most intense current concentrations are formed away from the magnetic nulls. The bowing mapping of current concentrations seen in Figure 2 in the lower right quadrant correlates well with a QSL, predicted by a large squashing factor (see Figure 1). The other current concentration site, which maps down to the rotating (left) polarity is, however, less well expressed in the squashing analysis. The corresponding QSLs mapping down to the three uppermost and left-hand-side quadrants of the area of interest. They do not create currents because they are not energized by photospheric plasma motion. Hence, creation and growth of coronal currents directly reflects the action of the energy input through the photosphere. Our results also demonstrate another limit of the QSL method in predicting current concentrations: If one compares the location of QSL (Figure 1) with the current concentrations, shown in Figure 2, one can see that the currents evolve over much broader areas. QSL, therefore, predict a skeleton around which currents may arise, if energy flows into the region.

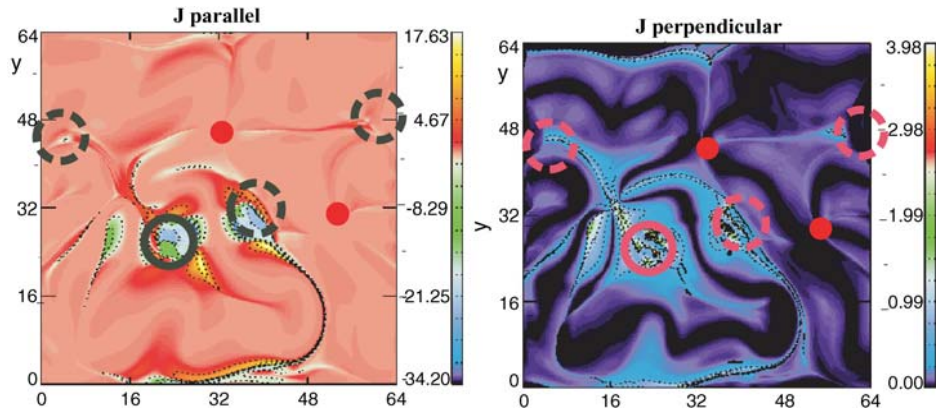

*Figure 2.* Simulated parallel (*left panel*) and perpendicular (*right panel*) currents, integrated along the magnetic field lines. For the meaning of the *red dots* and *solid*, respectively, and dashed circles see the caption of Figure 1. In the *right panel*, the *color* of the *circles* is changed to *black* for better visibility (from Büchner *et al.*, 2006).

## 4. Summary and Conclusions

Looking for the relation between QSL and current concentrations in the solar atmosphere, we simulated the response of a corona with only chromospheric magnetic nulls to an observed rotation of a photospheric flux concentration near an EUV-BP (Brown *et al.*, 2001). Comparing the photospheric footprints of parallel and perpendicular coronal current concentrations with the locations of large squashing factors, indicating strong changes of the magnetic connectivity, we found that the current sheet location correlates with some of the QSLs, obtained for the initial, force-free extrapolated observed photospheric longitudinal magnetic field. Both squashing factors, $Q$, based on the photospheric normal magnetic field components and $Q_1$ with the determinant consisting of derivatives of the Jacobian of the photospheric shift vectors in the denominator, show similar results.

The strongest currents are generated, however, not in regions of the largest squashing factors, i.e., near separatrices, but at locations, where two conditions are fulfilled simultaneously: large squashing and local energy inflow from the photosphere. Hence, one has always to combine the calculation of squashing factors with some information about the photospheric energy input in order to predict coronal current concentrations.

A technical caveat of searching for potential current sheet locations by means of squashing factors is the huge numerical effort: for each position of interest, one has to calculate three forward mappings to the conjugate photospheric footpoints and three backward-directed ones.

Obviously, there is room to look for further indicators and predictors of coronal current concentrations which might be simpler to obtain. One possible candidate is the differential flux tube volume (see, e.g., Büchner, 2005), or other indicators, discussed, e.g., by Demoulin *et al.* (1997) (who called the differential flux tube volume "delay function"). Note that the QSL concept, though it predicts some skeleton around which current sheets potentially might arise, does not include any ideas about the physical mechanism of current generation. Hence, like QSLs, separatrices, which can also be found by calculating the flux tube squashing, provide useful hints at current concentrations and, possibly, reconnection. However, it is still necessary to look for the physical process of current generation by energy input into the corona to predict the location of current sheets and, possibly, the sites of magnetic reconnection.

## Acknowledgements

The author thanks B. Nikutowski at the MPS Lindau for numerical calculations, A. Otto from the University of Alaska in Fairbanks for discussion and collaboration on developing the numerical code, P. Démoulin from the Observatoire Meudon and V. Titov from the SAIC, San Diego for exciting discussions of this topic, as well as an anonymous referee for carefully reading the manuscript and comments, which helped to improve the paper. A special thanks goes also to the German Research Foundation (DFG) for its support of project BU 777/2.